\documentclass[twocolumn,superscriptaddress,floatfix,prb,showpacs]{revtex4-1}
\usepackage[colorlinks=true]{hyperref}
\usepackage{graphicx}
\usepackage{subfig}
\usepackage{times}
\usepackage{amsmath}

\newcommand*{\citen}[1]{%
  \begingroup
    \romannumeral-`\x % remove space at the beginning of \setcitestyle
    \setcitestyle{numbers}%
    \cite{#1}%
  \endgroup
}
\begin{document}
\title{Evolution of the electronic and lattice structure with carrier injection in BiFeO$_3$}
\author{Xu He}
\affiliation{Beijing National Laboratory for Condensed Matter Physics, Institute of Physics, Chinese Academy of Sciences, Beijing 100190, China}
\author{Kui-juan Jin}
\email[email: ]{kjjin@iphy.ac.cn}
\affiliation{Beijing National Laboratory for Condensed Matter Physics, Institute of Physics, Chinese Academy of Sciences, Beijing 100190, China}
\affiliation{Collaborative Innovation Center of Quantum Matter, Beijing 100190, China}
\author{Hai-zhong Guo}
\affiliation{Beijing National Laboratory for Condensed Matter Physics, Institute of Physics, Chinese Academy of Sciences, Beijing 100190, China}
\author{Chen Ge}
\affiliation{Beijing National Laboratory for Condensed Matter Physics, Institute of Physics, Chinese Academy of Sciences, Beijing 100190, China}

\begin{abstract}
We report a density functional study on the evolution of the electronic and lattice structure in BiFeO$_3$ with injected electrons and holes. First, the self-trapping of electrons and holes were investigated. We found that the injected electrons tend to be localized on Fe sites due to the local lattice expansion, the on-site Coulomb interaction of Fe $3d$ electrons, and the antiferromagnetic order in BiFeO$_3$. The injected holes tend to be delocalized if the on-site Coulomb interaction of O $2p$ is weak (in other words, $U_\mathrm{O}$ is small). Single center polarons and multi-center polarons are formed with large and intermediate $U_\mathrm{O}$, respectively. With intermediate $U_\mathrm{O}$, multi-center polarons can be formed. We also studied the lattice distortion with the injection of carriers by assuming the delocalization of these carriers. We found that the ferroelectric off-centering of BiFeO$_3$ increases with the concentration of the electrons injected and decreases with  that of the holes injected. It was also found that a structural phase transition from $R3c$ to the non-ferroelectric $Pbnm$ occurs, with the hole concentration over 8.7$\times10^{19} cm^{-3}$. The change of the off-centering is mainly due to the change of the lattice volume. The understanding of the carrier localization mechanism can help to optimize the functionality of ferroelectric diodes and the ferroelectric photovoltage devices, while the understanding of the evolution of the lattice with carriers can help tuning the ferroelectric properties by the carriers in BiFeO$_3$.

\end{abstract}
\maketitle
\section{Introduction}
In transitional metal oxide perovskites, there is strong correlation between the degrees of freedom of charge and lattice. When extra charges are injected into those materials, they interacts with the lattice, causing novel phenomenon. Unlike in the conventional insulators and semiconductors, the change in BiFeO$_3$ (BFO) with injected carriers cannot be seen as a mere rigid shift of the band. The lattice distorts with the carrier injection, and the injected carriers can be trapped due to the lattice distortion. Here we investigate the behaviors of the injected carriers and the lattice in BFO with first principle methods.

BFO has been of great interests for many years\cite{ADMA:ADMA200802849}, because its large ferroelectric polarization and relatively small band gap\cite{PhysRevB.77.014110,Gujar2007142,IhlefeldAPL2008,Xu20083600,Clark2716868} make it a good choice as semiconductor and optoelectronic material\cite{basu2008photoconductivity,wang2013electro} in devices such as ferroelectric diode\cite{choi2009switchable,wang2011switchable,ge2011numerical} and ferroelectric photovoltaic device~\cite{yang2010above,kreisel2012photoferroelectric}. In these devices, carriers are injected into BFO either by electric field or optical excitation. A most important issue about the carriers is whether they tend to be localized or delocalized, as this greatly affects the mobility and the lifetime of the carriers and the leakage current in BFO. Therefore, the understanding of the carrier behavior in BFO is crucial for revealing the mechanisms behind its abundant properties, as well as for the development of the devices.

 There are a few evidences showing that the carrier has the tendency to be trapped in BFO. The electronic conductivity in non-doped and $p$-type BFO follows the $\log\sigma\propto 1/T$ law, implying the polaron hopping mechanism\cite{ADMA:ADMA200800218,doi:10.1021/cm300683e,0022-3727-42-16-162001}. Hole doping were achieved by substituting Bi or Fe ions with acceptor cations (like Ca$^{2+}$, Sr$^{2+}$, Ba$^{2+}$, Ni$^{2+}$, and Mg$^{2+}$)~\cite{doi:10.1021/cm300683e,wang2006effect,yang2009electric,QiBFOdoping}.Large concentration of acceptor cations tends to break the symmetry of bulk. For example, by substituting about 10\% Bi ions with Ca ions, there is a monoclinic to tetragonal phase transition in BFO thin films\cite{yang2009electric}. Whereas it's difficult to achieve $n$-type doping; substituting Fe ions with Ti$^{4+}$ or Nb$^{5+}$ decreases the conductivity in BFO~\cite{QiBFOdoping, Jun2005133}. In the chemically doped BFO structures, whether the polarons are bounded to dopants or self-trapped is not clear. Schick \emph{et al.} studied the dynamics of the stress in BFO due to the excited charge carriers with ultrafast X-ray diffraction and found that the carriers tend to be localized\cite{PhysRevLett.112.097602}. Yamada \emph{et al.} found photocarriers can be trapped by means of transient absorption and photocurrent measurements\cite{yamada2014measurement}.The trapping of the carriers can happen because of the defects or the self-trapping effect in BFO. In the latter case, the carriers reduce their energies due to the local lattice distortion and form small polarons. The states of the trapped carriers are in the band gap, thus these carriers need energy to be excited and become conducting. In-gap states were observed in absorption spectra and photoluminescence measurements\cite{PhysRevB.79.134425,pisarev2009charge,PhysRevB.79.224106,hauser2008characterization}, while whether these states should be attributed to defect states or self-trapped states has not been clear yet. There has been extensive study on the defects states~\cite{ederer2005influence,PhysRevB.85.104409,zhang2010density}, whereas the study into the self-trapped state is lacking. In this work, we firstly investigate the self-trapping of the injected electrons and found that the electrons tend to be localized even when the defects are absent. The localization of injected holes were also studied. We found that the holes tend to be delocalized, to form multi-centered polarons, and to form single centered polarons if the on-site Coulomb interaction of O $2p$ electrons is weak, intermediate, and strong, respectively. The lattice distortions near the localized electrons/holes were also studied.

Another important issue is that how the lattice deforms if the injection of carriers are delocalized. The injected carriers, which are affected by the lattice, affect the lattice in return, thus can modulate the ferroelectric distortions. In ferroelectrics, the off-centering of ions, which is stabilized by the long-range Coulomb interaction, tend to be unstable with free charge, as the free carriers can screen the Coulomb interaction. However, ferroelectric metal, in which ferroelectric displacement coexists with conducting carriers, was predicted by Anderson and Blount~\cite{anderson1965symmetry} and then identified in LiOsO$_3$\cite{shi2013ferroelectric}. In some ferroelectrics, the ferroelectric displacement can survive within a range of carrier concentration.  For example, BaTiO$_3$, another ferroelectric perovskite, undergoes a phase transition from ferroelectric tetragonal phase to cubic with the injection of electrons above a critical concentration\cite{iwazaki2012doping,wang2012ferroelectric}. Can the ferroelectricity of BFO sustain the carrier injection? If it can, how is the ferroelectric displacement tuned by charges? In this work, we also studied the evolution of the lattice structure with the injection of carriers. We found that a structural phase transition from $R3c$ to the non-ferroelectric $Pbnm$ structure occurs, if the hole concentration is over a criterion of $8.7\times10^{19} cm^{-3}$. This indicates that hole injection can be used as an efficient way of depolarization of BFO if holes tend to be delocalized. Whereas the free electrons do not destabilize the ferroelectric distortion, but enhance the structural off-centering of BFO, which supports the idea that long-range ferroelectric order can be driven by short-range interactions\cite{PhysRevB.90.094108}.

\section{Methods}
The density functional theory (DFT) calculations have been performed using the local spin density approximation\cite{perdew1981self} (LSDA) and projector augmented wave method\cite{kresse1999ultrasoft} as implement in the Vienna \emph{ab initio} simulation package (VASP)\cite{kresse1996efficient}. A plane-wave basis set with the energy cutoff of 450 eV were used to represent the wave functions.

The localization of the carrier depends on whether the localized electronic state can form within the band gap. Therefor a good description of the band gap is needed. Local density approximation (LDA) and generalized gradient approximation (GGA) calculations always underestimate the band gap and tend to fail in predicting the localization of carriers. Our LDA calculation gives a band gap of 0.5 eV while the experimental band gap of BFO is about 2.8 eV. The  DFT +\emph{U} method can improve the description of the electronic properties in BFO~\cite{neaton2005first} by adding a Hubbard \emph{U}~\cite{liechtenstein1995density,dudarev1998electron} correction. Goffinet \emph{et al.}~\cite{goffinet2009hybrid} compared the results of DFT+\emph{U} and hybrid functionals and found that both can describe the structural properties well. The band gap with the hybrid functional B1-WC calculation is 3.0 eV, while the LDA+\emph{U} calculation with a $U_\mathrm{Fe}=3.8$ eV gives a 2.0 eV band gap. We used the more computationally inexpensive LDA+\emph{U} correction in all our calculations. An effective $U_\mathrm{Fe} = 4$ eV, which can give qualitative and sub-quantitative correct result for the structural, magnetic, and electronic properties in BFO, is used throughout this paper unless otherwise stated. In the calculations of the hole polarons, various $U_\mathrm{O}$'s ranging from 0 eV to 12 eV were used. Adding Hubbard $U$ to O $2p$ was found to be an effective way for the calculation of the hole polarons in titanite perovskites\cite{PhysRevB.90.035204}, in which the valence band maximum (VBM) is mostly O $2p$ states.

Bulk BFO adopts the symmetry with space group $R3c$, which can be viewed as pseudo cubic structure with a ferroelectric polarization along the [111] direction. We constructed $\sqrt{2}\times\sqrt{2}\times 2$, $2\times 2\times 2$, and $2\sqrt{2}\times 2\sqrt{2}\times 2$ pseudo cubic supercells, and by adding/removing one electron from the supercells, the concentration of the electrons/holes in these supercells are 1/4, 1/8, and 1/16 u.c.$^{-1}$, respectively. The $\sqrt{2}\times\sqrt{2}\times 2$ and $2\sqrt{2}\times2\sqrt{2}\times 2$ supercells are constructed from the structure in $Cc$ phase. The structures in $Cc$ phase and $R3c$ phase are very close. If the structure of the two phases are both put in  $2\times 2\times 2$ supercells, the only difference between them would be that the angles ($\alpha, \beta$ and $\gamma$) of the lattice parameters for the $R3c$ phase are all about 89.9$^\circ$, while $\alpha$ and $\gamma$ are fixed to 90$^\circ$ in the $Cc$ phase. The localized electrons/holes break the symmetry of the bulk. Here, a $5\times5\times5$ $\Gamma$-centered k-point grid were used to integrate the Brillouin Zone.
%We also carried out an calculation with a $2\times 2\times 2$ k-point mesh and the $4\times 4\times 4$ supercell (i.e. the carrier concentration is 1/64 u.c.$^{-1}$) to check whether self-trapping still happens when the carrier concentration is small.
G-type antiferromagnetic structure was assumed in all the calculations. The image charge correction\cite{PhysRevB.51.4014} and the potential-alignment correction\cite{PhysRevB.78.235104} were utilized in the DFT calculations with the adding and removing of the electrons.

To find whether the electron injected into the BFO is delocalized or localized, we compared the two states with and without the bulk symmetry being broken. By following the recipes of Deskins \emph{et al.}\ ~\cite{doi:10.1021/jp2001139},we first elongate the Fe-O bonds around one Fe site to break the transition symmetry. Then we set the initial magnetic moment of the specific Fe site 1 less than those of the other Fe sites, since Fe$^{3+}$ ion has the $3d^5$ high spin electronic configuration, adding one electron will reduce the net magnetic moment.  By using this as the initial state and relaxing the structure, the localized polaronic state can be obtained if there is a localized state within the band gap of BFO. Similar method can be applied in the calculation related to the hole localization. The initial structures were constructed by stretching or compressing the bonds near the hole center. In BFO without injected holes, O $2p$ states are almost fully occupied, thus have 0 spin. An 1 $\mu_B$ magnetization was set as the initial value for the O ion where the hole is assumed to be localized.

 To see how the lattice distorts with the carrier concentration, the symmetries of the lattices are fixed to a few low energy phases, namely the $R3c$, $Cc$, $R\bar{3}c$ , $Pbnm$, and $Pbn2_1$, respectively. A $5\times5\times5$ $\Gamma$-centered k-point mesh was used with these calculations.
\section{Results and Discussion}
%\subsection{electron injection}
\subsection{Bulk properties}
\begin{figure}[htbp]
  \centering
  \includegraphics[width=0.46\textwidth]{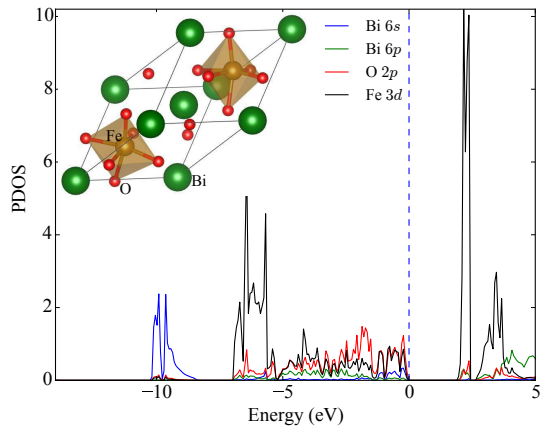}
  \caption{The density of Bi $6s$ and $6p$, Fe $3d$, and O $2p$ states. The inset shows the primitive cell of BFO with $R3c$ symmetry. The results were calculated with $U_\mathrm{Fe}$=4 eV and $U_\mathrm{O}$=0 eV. Changing the values of $U_\mathrm{Fe}$ and $U_\mathrm{O}$ does not change the nature of the VBM and CBM.\label{fig:LLDOS} }
\end{figure}

 Here we look into the bulk properties of BFO. The primitive cell of BFO with $R3c$ symmetry is shown in the inset of Fig.~\ref{fig:LLDOS}. There are 10 atoms in the primitive cell, including two 2 Bi atoms, 2 Fe atoms, and 6 O atoms. Each Bi atom has 12 neighboring O atoms; each Fe atom has 6 neighboring O atoms which make an octahedron. The calculated structural parameters  with various $U$'s are given in table~\ref{tab:structure}, which agree well with experimental data\cite{Kubel:du0188} and previous calculations(e.g. in Ref.~\citen{neaton2005first}).

The partial density of states of BFO is in Fig.~\ref{fig:LLDOS}. The states at the conduction band minimum (CBM) are mostly the Fe $3d$ states. Consequently, the injected electrons mainly stay at the Fe sites. The valence band maximum (VBM) consists of O $2p$, Fe $3d$, and Bi $6s$ states. Though the Bi $6s$ states are deep below the Fermi energy, the strong hybridization between the Bi 6s and O 2p orbitals lead to considerable Bi $6s$ DOS at the VBM. The electron lone pair, which is the driving force of the ferroelectricity in BFO, is related to the Bi 6s-O $2p$ antibonding states at the VBM~\cite{PhysRevB.74.224412}. Which site are the injected holes localized (if they tend to be localized) at needs to be investigated.

\begin{table}[htbp]
\caption{The structural parameters of BFO with $R3c$ symmetry calculated with various $U$'s. The Wyckoff positions are Bi 2a $(x, x, x)$, Fe 2a $(x, x, x)$, and O 6b $(x, y, z)$.\label{tab:structure}}.
\begin{center}
\begin{tabular}{l c c c c c c}
\hline
\hline
 ($U_\mathrm{Fe}$, $U_\mathrm{O}$) & & (4, 0)\footnote{Result from LDA+U calculation, this work\label{thiswork}.}& (4, 0)\footnote{Result from LDA+U calculation, Ref.~\citen{neaton2005first}.}&(4, 8)\textsuperscript{\ref{thiswork}}&(4, 12)\textsuperscript{\ref{thiswork}}& Exp\footnote{Experimental result, Ref.~\citen{Kubel:du0188}.}\\
\hline

  Bi (2a)&$x$ &0 &0 &0 &0 &0 \\
  Fe (2a)&$x$ &0.226 &0.227 &0.227&0.227 &0.221 \\
  O  (6b) & $x$ &0.540 &0.542 &0.538&0.536 &0.538 \\
  & $y$ &0.942 &0.943 &0.942&0.940 &0.933\\
  & $z$ &0.397 &0.397 &0.398&0.399 &0.395\\
  $a_{rh}$ (\AA)& &5.52 &5.52 &5.49 &5.47 &5.63 \\
  $\alpha$ (deg)& &59.79 &59.84 &59.82 &59.72 &59.35 \\
\hline
\hline
\end{tabular}
\end{center}
\end{table}
\subsection{Self trapping of electrons}
\begin{figure}[htbp]%
     \includegraphics[width=0.45\textwidth]{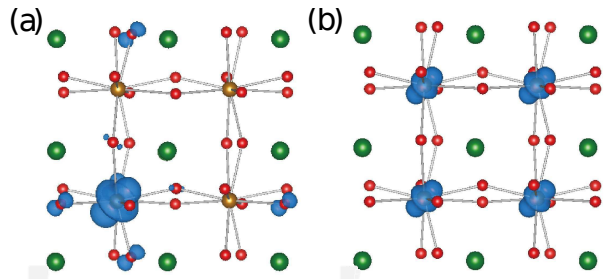}
    \caption{The isosurface of the (a) localized (b) delocalized charge corresponding to the density of 1/8 e$^-$/u.c. The green, brown, and red spheres represent the Bi, Fe, and O ions, respectively. }
    \label{fig:isosurface}%
\end{figure}

Electrons injected into the BFO lattice can be either delocalized or localized, depending on how they interacts with the lattice. The delocalized electrons stay on the CBM and the symmetry of the lattice is preserved. Whereas the localized electrons break the symmetry of the lattice, and change the local chemical bonds to lower the energy, forming an in-gap state, i.e. forming a small polaron. To understand the behavior of the injected electrons, we compared the two kinds of electron states with the DFT+$U$ calculations. Figures~\ref{fig:isosurface} (a) and (b) show the electron density isosurfaces for the localized and the delocalized state, respectively. The localized electron resides mostly on one Fe site and the delocalized electron distributes on all Fe sites.

\begin{figure}[htbp]
  \centering
    \includegraphics[width=0.45\textwidth]{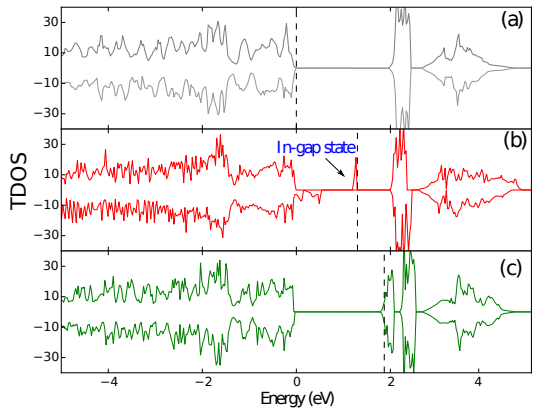}
  \caption{The total density of states of  BFO in the $2\times 2\times 2$ supercell (a) without injected electron (b) with one localized injected electron. (c) with one injected delocalized electron. The states with the energies below the dashed line are occupied.   \label{fig:TDOS}}
\end{figure}

To see whether the in-gap state is stable, we calculated the electronic structures of the BFO with localized and delocalized injected electron. The total density of states (TDOS) of the $2\times 2 \times 2$ supercell are shown in Fig.~\ref{fig:TDOS}. The TDOS of BFO without injected electrons is used as a reference. In the localized case, there is a in-gap state of about 0.6 eV below the CBM, which corresponds to the localized electron state. The in-gap state are 0.5 and 0.7 eV below the CBM in the supercells with electron concentration of 1/4 and 1/16 u.c.$^{-1}$, respectively. As for the delocalized state, the change is not just a Fermi energy shift within rigid bands, either. A split in the formerly unoccupied Fe $t_{2g}$ band can be clearly seen. The possible reasons are the change in the lattice and the electron-electron interaction, which shift the occupied bands down and unoccupied bands up.

The electron self-trapping energy $E_{EST}$ defined as
\begin{equation*}
  \label{eq:Est}
  E_{EST}=E_{tot}(BFO:e_{CBM}^-)-E_{tot}(BFO:e_{polaron}^-) ,
\end{equation*}
where $E_{tot}(BFO:e_{CBM}^-)$ is the total energy of BFO cell with an injected electron at the CBM, $E_{tot}(BFO:e_{polaron}^-)$ is the total energy of the BFO with a localized electron. A positive value means that the small polaronic state is energetically preferable. The $E_{EST}$ of the supercells with electron concentration of 1/4, 1/8, and 1/16 u.c.$^{-1}$ are 0.66, 0.50, and 0.39 eV, respectively.

We analyzed the possible mechanism for the self-trapping of electrons, and found that the self-trapping is driven by the local lattice expansion and the Coulomb repulsion of the Fe 3$d$ electrons, and is stabilized by the antiferromagnetic structure.

\begin{figure}[htbp]
  \centering
    \includegraphics[width=0.48\textwidth]{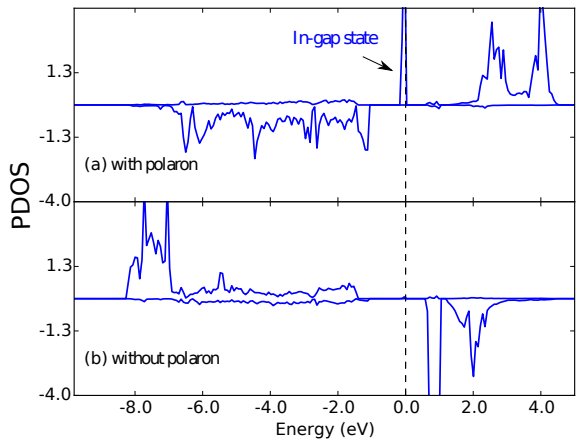}
  \caption{The projected density of the $3d$ states of (a) Fe site with a injected localized electron (b) neighboring Fe site. The energies are shifted so that the states below the energy 0 eV are occupied. \label{FeDOS}}
\end{figure}

One reason for the self-trapping of electrons is the distortion of the lattice surrounding the electrons. The most obvious change of the lattice is the expansion of the Fe-O octahedra where the injected electrons are localized. In the ferroelectric BFO, the six Fe-O bonds of each Fe ion can be divided into two groups, the group of longer bonds and the group of shorter bonds, as the Fe atoms do not reside at the center of the oxygen octahedra. We found both groups of Fe-O bonds near the injected electrons are elongated as listed in table~\ref{tab:bondlengths}. The elongation of the bonds can be easily understood as the consequences of the Coulomb repulsion between the injected electron and the negatively charged oxygen ions. Because of elongation of the Fe-O bonds, the Coulomb energy of the injected electrons is reduced. Meanwhile, this elongation reduces the Fe $3d$-O $2p$ overlap, suppressing the hopping of the injected electrons and increasing the tendency of localization.

The higher the carrier concentration is, the larger the elongation is, as the elongation of the Fe-O bonds rises the elastic energy of the surrounding lattices. The difference of the energy between the localized state and the CBM is larger in the structure with higher electron concentration, which is consistent with the longer local Fe-O bonds as shown in Table~\ref{tab:bondlengths}. On the other hand, the E$_{EST}$ is smaller in the structure with lower electron concentration because of the increasing of the elastic energy cost. %For the structure with electron concentration of 1/16  u.c.$^{-1}$, the Fe-O bond lengths without localized electron on the Fe site are almost identical with the corresponding bulk values without injection, which means the strain caused by the expansion of the octahedron is relaxed. Thus, further reducing the electron concentration below 1/16  u.c.$^{-1}$ would no longer reduce the self-trapping energy.

\begin{table}[h]
\caption{The lengths of Fe-O bonds. In each Fe-O octahedron, the six Fe-O bonds can be divided into two groups of three long bonds and three short bonds, labeled by the subscript $L$ and $S$ respectively. The superscript $e$ means that a localized electron resides in the octahedron. The $l_L$ and $l_S$ are the lengths of bonds in the octahedron farthest away from the localized electron. All units are $\text{\AA}$. }
\label{tab:bondlengths}
\begin{center}
\begin{tabular}{c c c c c}
\hline\hline
 electron concentration & $l_L^e$ & $l_S^e$ & $l_{L}$ & $l_{S}$\\
\hline
 no injection & - & - & 2.05 & 1.94 \\
\hline
% 1/64 $e^{-}$/u.c. & 2.11 & 2.00 &2.06 &1.94\\
%\hline
1/16 $e^{-}$/u.c. & 2.11 & 2.01 & 2.05 & 1.93\\
\hline
1/8 $e^{-}$/u.c.& 2.12 & 2.01 & 2.07 & 1.94\\
\hline
1/4 $e^{-}$/u.c.& 2.15 & 2.04 & 2.11 & 1.94\\
\hline
\end{tabular}
\end{center}
\end{table}

% \begin{figure}[htb]
%   \centering
%     \includegraphics[width=0.45\textwidth]{figure/varec.eps}
%   \caption{The lengths of Fe-O bonds. The lengths of the longer and the shorter bonds are denoted as $l_L$ and $l_S$, respectively. The superscript e means there is a localized electron on the Fe site. All units are $\text{\AA}$. Because the symmetry of the structures with localized electrons are broken, not all bonds in the same group are equivalent. The values are within a error of $\pm$0.02.\footnote{This figure is the same as Table I. Just here to see which one is better, the table or the figure.} \label{fig:varelecconcentration}}
% \end{figure}

\begin{figure}[htbp]
  \centering
    \includegraphics[width=0.45\textwidth]{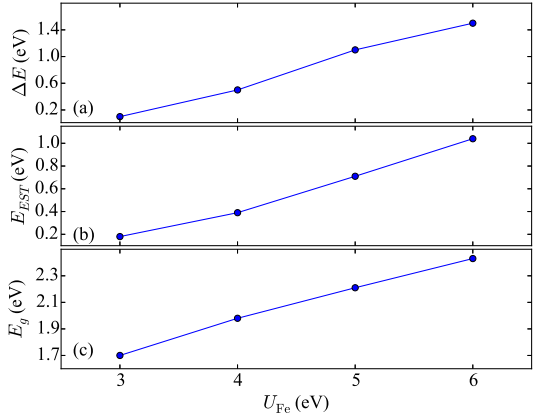}
  \caption{The dependencies of (a) $\Delta E$ (The difference between the energy of the in-gap state and that of the lowest unoccupied state) and (b) the self-trapping energy $E_{EST}$ (c) The calculated band gap on $U_{\mathrm{Fe}}$. \label{fig:varU} }
\end{figure}

Another reason for the self-trapping of the electrons is the Coulomb repulsion effect of the electrons. To see how this influences the localization of the electrons, we calculated the electronic structure with various effective $U_\mathrm{Fe}$'s ranging from 0 to 6 eV. The self-trapping happens only if $U_\mathrm{Fe}>$2 eV. We found the difference between the energy of the in-gap state and that of the lowest unoccupied state is larger with larger $U_\mathrm{Fe}$, as shown in Fig.~\ref{fig:varU} (a). The in-gap state and the lowest unoccupied state are both Fe $3d$, thus the former can be seen as the lower Hubbard band (LHB) and the later as the upper Hubbard band (UHB). The on-site Coulomb repulsion of the Fe $3d$ electrons shift the LHB down and the UHB up, enlarging the difference between them. Because the Coulomb repulsion lowers the energy of the localized electron, the self-trapping of the electrons are stabilized. Therefore, the self-trapping energy is higher with larger $U_\mathrm{Fe}$, as shown in Fig.~\ref{fig:varU} (b).

In all the structures, our calculations gave the results of G-type antiferromagnetic order with a total magnetic moment of 1 $\mu_B$ when one electron is self-trapped. The projected density of states of Fe $3d$ orbitals are shown in Fig.~\ref{FeDOS}. The in-gap state has the opposite spin with the other occupied $3d$ states on the same site. Therefore, the electronic configurations of the Fe ions are $d^5\downarrow d^1\uparrow$ and $d^5\uparrow$ with and without the localized electron, respectively (Figs.~\ref{FeDOS} (a) and (b)). In the Fe sites neighboring to that with localized electron, the five $3d$ states with the same spin as the in-gap state are fully occupied, which makes the hopping to the nearest neighbors forbidden. Therefore, the antiferromagnetic order stabilize the localization of the injected electron.

%BFO structure with hole injection is also studied with LDA+\emph{U} calculations. No localized hole state was found for a large range of \emph{U}'s and hole concentrations, which means the delocalized hole state is energetically preferable. We found that there is a phase transition from $R3c$ to $Pbnm$ if the concentration of hole is larger than $8.7\times10^{19} cm^{-3}$, which will discussed later in this paper.
% The total density of states of BFO in  these two phases are in Fig.~\ref{r3cvspbnm}. The electronic structures of these two phases are quite similar.
%\begin{figure}[htb]
%  \centering
%    \includegraphics[width=0.45\textwidth]{figure/r3cvspbnm.eps}
%  \caption{The density of states of hole injected BFO of (a) $R3c$ strucuture and (b) $Pbnm$ structure. The states below the energy 0 eV are occupied. \label{r3cvspbnm}}
%\end{figure}
\subsection{Self trapping of holes}
\begin{figure}[htbp]
  \centering
  \includegraphics[width=0.5\textwidth]{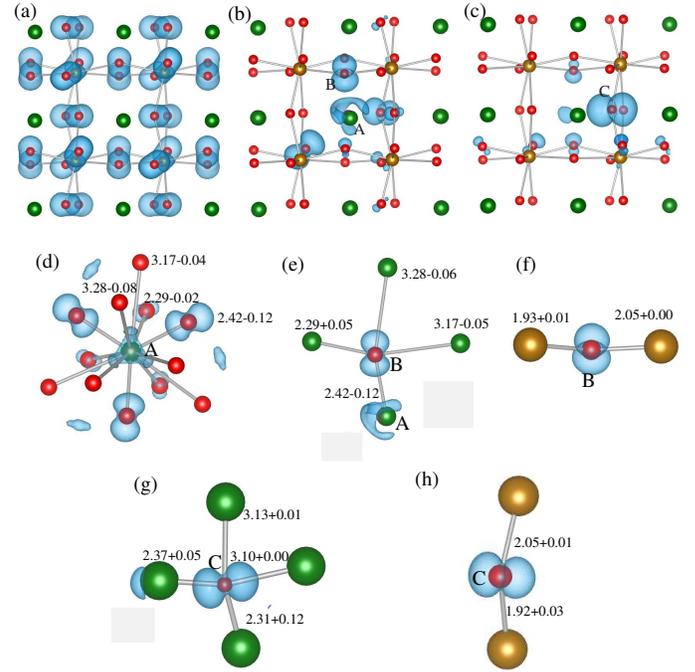}
  \caption{The isosurface of density of injected holes in the forms of (a) delocalized holes, (b) multi-center hole polarons, and (c) single-center small polarons in a $2\times 2\times 2$ supercell.  (d-f) and (g-h) show the local lattice distortion near the multi-center polaron and the single-center polaron, respectively. In (d), (e), and (g), only Bi-O bonds are shown. In (f) and (h),only Fe-O bonds are shown. The bond lengths are written in the form of $l_o+\Delta l$, where $l_0$ is the bond length in the bulk structure with no carrier injection, and $\Delta l$ is the increment. (b), (d), (e), and (f) were calculated with $U_\mathrm{O}$=8 eV. (c), (g) and (h) were calculated with  $U_\mathrm{O}$=12 eV. The green, brown, and red spheres represent the Bi, Fe, and O ions, respectively.\label{fig:holepolaron}}
\end{figure}

\begin{figure}[htbp]
\includegraphics[width=0.47\textwidth]{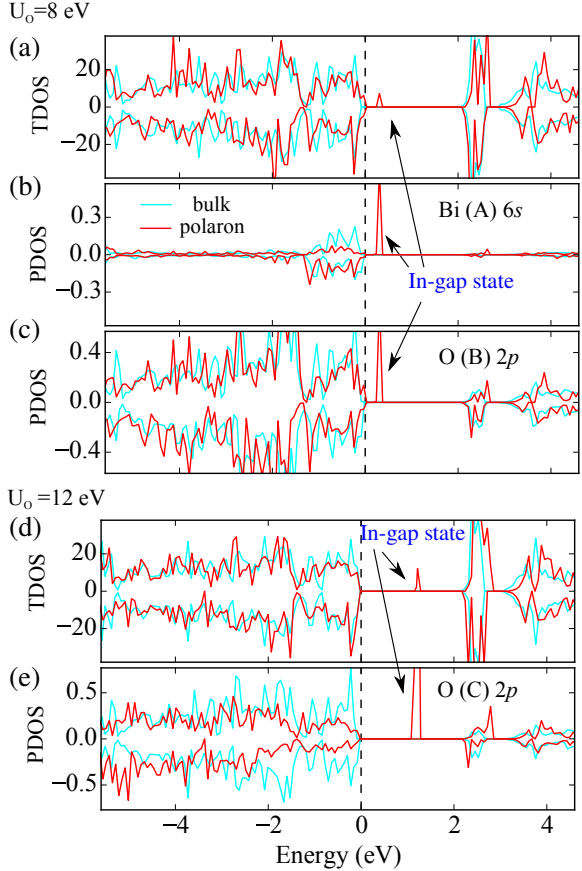}
\caption{(a) Total density of states of BFO in the $2\times 2\times 2$ supercell with one hole polaron. (b) Partial density of states of the Bi and O atoms with and without localized holes. (a) and (b) were calculated with $U_\mathrm{O}$=8 eV.  (c) Total density of states of BFO with delocalized and localized holes. (d) Partial density of states of the O atoms with and without localized holes. (c) and (d) were calcuated with $U_\mathrm{O}$=12 eV. The red curves are the DOSes of the structure with one hole polaron. The cyan curves are the DOSes of the BFO without hole injection and are plotted as reference.\label{fig:holeDOS}}
\end{figure}

The self-trapping of holes was also investigated. In BFO, the top of the valence band is a mix of O $2p$, the Fe $3d$ and Bi $6s$ states, as shown in Fig.~\ref{fig:LLDOS}. Therefore, we need to know what sites would the polarons reside on if the holes are self-trapped. We found that Fe-site centered small polaron is energetically unfavorable with a large range of $U_\mathrm{Fe}$ from 0 eV to 8 eV. % The effect of the Hubbard $U$ is maximal when the orbitals are half occupied. In BiFeO$_3$, five of the ten Fe $3d$ states are occupied. Therefore, the hole state on the Fe site, which change the Fe $3d^5$ configuration to $3d^4$, would rise the energy level of the occupied Fe $3d$ states and is thus unfavorable.
The largest contribution to the top of valence band is from the O $2p$ states. We explored the self-trapping of holes by adding Hubbard $U$ to the O $2p$ states~\cite{PhysRevB.90.035204}. Using various $U_\mathrm{O}$ from 0 to 12 eV, we found that holes tend to be delocalized with $U_\mathrm{O} <$ 6 eV; small polarons centered on O sites are stabilized for $U_\mathrm{O} \geq$ 12 eV; multicenter polarons are formed if $U_\mathrm{O}$ is between 6 eV and 12 eV. The delocalized holes, multi-center polaron, and the single-center small polaron are shown in Fig.~\ref{fig:holepolaron} (a), (b), and (c), respectively. The delocalized hole mainly distributes uniformly at the O sites; the multi-center hole polaron stays on the hybridized orbital of Bi and O (mainly at three O sites and the Bi site near their center); the single center hole polaron mainly stays at one O site. The TDOS calculated with $U_\mathrm{O}$=8 eV and $U_\mathrm{O}$=12 eV are in Fig.~\ref{fig:holeDOS} (a) and (d), respectively. States inside the band gap of the structure emerges, which corresponds to the polarons.

\begin{figure}[htbp]
  \centering
  \includegraphics[width=0.46\textwidth]{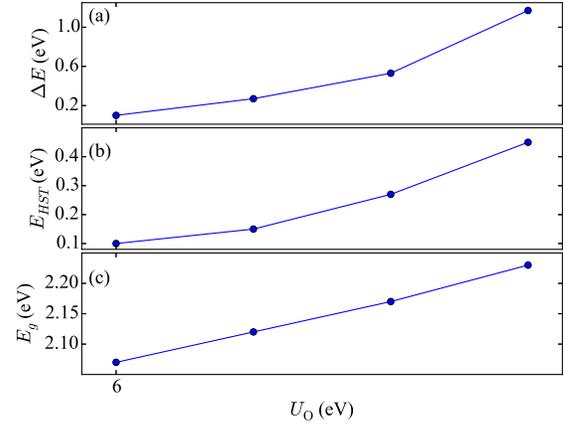}
  \caption{\label{fig:varUO} The dependences of (a) the energy difference between the in-gap state and the valence band maximum, (b) the hole self-trapping energy, and (c) the band gap of BFO on $U_\mathrm{O}$}
\end{figure}

 The change of the band gap ($E_g$) with $U_\mathrm{O}$ is small since the O bands are almost fully filled in bulk BFO (Fig. ~\ref{fig:varUO} (c)). Adding $U$ to the O $2p$ state does not significantly improve the lattice structure results, as can be seen from table~\ref{tab:structure}. Therefore, we do not intend to acclaim what value of $U_\mathrm{O}$ is most appropriate for describing the self-trapping of the holes. Thus we also don't acclaim whether and what kind of hole polarons tend to be formed here. Instead, we study the properties of the polarons by varying the $U_\mathrm{O}$.

Like the self-trapping of electrons, the self-trapping of holes is also stabilized by on-site electron Coulomb interaction.  Since most of the hole states are O $2p$, the dependence on $U_\mathrm{Fe}$ is not significant. We studied the dependence of the self-trapping energy and the in-gap state energy on $U_\mathrm{O}$. The hole self-trapping energy $E_{HST}$ defined as
\begin{equation*}
  \label{eq:Hst}
  E_{HST}=E_{tot}(BFO:h_{VBM}^+)-E_{tot}(BFO:h_{polaron}^+) ,
\end{equation*}
where $E_{tot}(BFO:h_{VBM}^+)$ is the total energy of BFO supercell with an injected hole at the VBM, $E_{tot}(BFO:h_{polaron}^-)$ is the total energy of the BFO supercell with a hole polaron. The $U_\mathrm{O}$ dependence of the $E_{HST}$ was studied. The $E_{EST}$ increases with $U_\mathrm{O}$, as shown in Fig.~\ref{fig:varUO} (b). In BFO without carrier injection, the O $2p$ states are almost fully occupied. With the removing of one electron, the in-gap state and the occupied O $2p$ states can be seen as the UHB and LHB of the O $2p$, respectively. The effect of $U_\mathrm{O}$ is to push the in-gap state (the UHB) up and the occupied O 2p states (the LHB) down, which lowers the total energy. Figure~\ref{fig:varUO} (a) shows that the energy difference ($\Delta E$) between the in-gap state and the VBM increases with $U_\mathrm{O}$, which is consistent with the larger UHB/LHB splitting.

The multi-center hole polaron state is a mix of Bi $6s$ and O $2p$ state, indicating that the hybridization between them is strong and plays an important role. The PDOSes of Bi $6s$ and O $2p$ on the sites corresponding to the ions marked as A and B in Fig.~\ref{fig:holepolaron} (b) are shown in Fig.~\ref{fig:holeDOS} (b) and (c). It can be seen that both the Bi $6s$ and O $2p$ state components are in the in-gap state. Instead of pushing one O $2p$ orbital up into the band gap, the Hubbard $U$ on O $2p$ pushes the hybridized (Bi $6s$, O $2p$) state up. The delocalization effect of the hybridization competes with the localization effect of the on-site electron Coulomb interaction. With small $U_\mathrm{O}$ ($<$6 eV), the delocalization is predominant, leading to free holes. With large $U_\mathrm{O}$ ($>$12 eV), the localization becomes predominant, leading to single-center small polarons. With intermediate $U_\mathrm{O}$, multi-center polarons are formed.

 Here we look into the local lattice distortion near the multi-center polaron. The multi-center polaron does not break the 3-fold rotation symmetry. The rotation axis is along [111] and through the Bi ion marked as A in Fig.~\ref{fig:holepolaron} (b). The lengths of the bonds between this Bi ion and O ions decreases as the polaron is formed (Fig.~\ref{fig:holepolaron} (c)). Since the Bi $6s$ and O $2p$ states are antibonding at the top of valence band, the decreasing of the Bi-O bond length enhance the Bi $6s$ - O $2p$ hybridization and further pushes the unoccupied anti-bonding state up. Consequently, the in-gap state is stabilized by the lattice distortion. The change to the lengths of Fe-O bonds is relatively small (Fig.~\ref{fig:holepolaron} (e)).

The single-center hole polaron is mostly on one O $2p$ orbital as shown by the spatial distribution of the hole (Fig.~\ref{fig:holepolaron} (c)) and the PDOS (Fig.~\ref{fig:holeDOS} (e)). For the single-center polaron state, the on-site energy play a more important role than the inter-site orbital hybridization. The lengths of the Bi-O bonds and Fe-O bonds for the O site where the hole is localized increase (Figs.~\ref{fig:holepolaron} (g) and (h)), i.e. the distances between the hole and the positively charged ions increase. Thus the Coulomb energy is reduced, which stabilizes the self-trapping of holes on the O site.

\subsection{Lattice deformation with delocalized carriers }
\begin{figure}[htbp]
  \centering
  \includegraphics[width=0.45\textwidth]{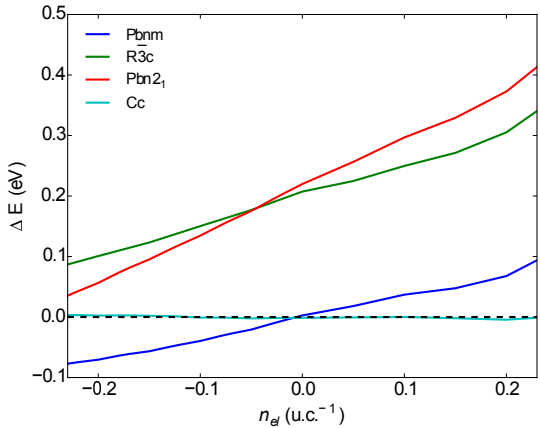}
  \caption{Calculated total energy difference versus injected carrier concentration with $R3c$ structure in various possible structural arrangements. The dashed line at 0 eV denote the energy of the $R3c$ structure.}
  \label{fig:energy}
\end{figure}

Here we investigate the distortion of the lattice under the assumption that the injected carriers are delocalized. We calculated the total energy of various structural arrangements ($R3c$, $Cc$, $R\bar{3}c$, $Pbnm$, $Pbn2_1$) with the change of concentration of delocalized carriers. The structure of the $Cc$ phase is very close to the $R3c$ structure. Therefore, the energy difference between the $R3c$ and $Cc$ phase is almost zero and we do not distinguish these two phases here. The $R\bar{3}c$ is the paraelectric phase of BFO at high temperature. The $Pbnm$ phase is featured with antiferroelectric oxygen octahedron rotations, which compete with the ferroelectric distortion. In the $Pbn2_1$ stucture, the antiferroelectric oxygen octahedron rotations coexist with Bi ion off-centering displacements. The results are shown in Fig.~\ref{fig:energy}. The $R3c$ structure is energetically preferable with electron injection. For hole concentration larger than 0.005 hole per BFO unit (about 8.7$\times 10^{19} cm^{-3}$), the orthorhombic $Pbnm$ structure, which is not ferroelectric, is energetically preferable. Therefore, BFO of $R3c$ tends to be depolarized with hole injection. The estimated value of the critical hole concentration is quite rough, as the energy difference between the phases near the phase transition point is small. It also depends on the functional used. The Perdew-Burke-Ernzerhof~\cite{PBE} (PBE) functional plus $U$ with $U=4$ gives a concentration of about 0.08 u.c.$^{-1}$ (about $6.23\times10^{21}$ cm$^{-3}$). But the trend toward the phase transition is robust. Neither 0.005 nor 0.08 holes per unit cell is a too large number, which indicates that hole injection can be an efficient way to depolarize the BFO if the holes are delocalized.

\begin{figure}[htbp]
  \centering
  \includegraphics[width=0.45\textwidth]{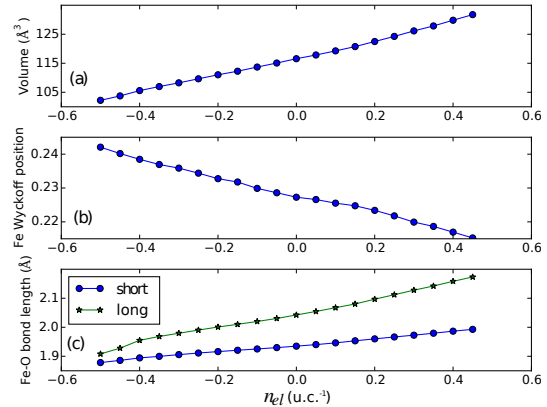}
  \caption{The change of lattice with adding/removing delocalized electrons. (a) The volume of the BFO $R3c$ primitive cell (two BiFeO$_3$ formula units). (b) The Wyckoff position of Fe. The central symmetry Wyckoff position of Fe is 0.25. (c) The Fe-O bond lengths.}
  \label{fig:latticedelocal}
\end{figure}

The details of the evolution of the lattice structure with the $R3c$ symmetry kept are shown in Fig.~\ref{fig:latticedelocal}. The volume of the lattice increases with electron injection, and decreases with hole injection, as shown in Fig.~\ref{fig:latticedelocal} (a). The absolute positions of the band edges
shift in order to minimize the electronic energy, which is achieved by
changing the volume. The ferroelectric off-centering of BFO has two main features, one being that the Fe site with Wyckoff position $2a (x,x,x)$ deviates from the centrosymmetric position $x = 0.25$, the other being that Fe-O bonds form two groups of longer and shorter bonds. The Wyckoff positions of Fe and the lengths of Fe-O bonds are shown in Fig.~\ref{fig:latticedelocal} (b) and (c), respectively. The off-centering is stronger as the concentration of the injected electrons increases. The trends are opposite with the injection of holes into the BFO structure. In summary, the injection of depolarized electrons enhances the off-centering of the $R3c$, whereas that of the holes reduces the off-centering.

The change of the lattice structure with concentration of carriers is very much alike to the change with the hydrostatic strain. A $R3c$ to $Pbnm$ transition with a hydrostatic pressure were predicted and found\cite{PhysRevB.74.224412,PhysRevB.79.184110}. Di\'eguez \emph{et al.}\cite{PhysRevB.83.094105} proposed that the reduction in structural off-centering and the phase transition are because of the less directional Bi-O bonds caused by the decreasing of the lattice volume. Just like in the hydrostatic compressed structures, the volume of the unit cell, the off-centering of the Fe cations, and the difference in the short and long Fe-O bonds are reduced in the hole injected structure, as shown in Fig.~\ref{fig:latticedelocal}.

Because of the similarity in the structural evolutions with carrier injection and hydrostatic pressure, the two kinds of evolutions can have the same origin. The reason for the weakening of the structural off-centering can be that the Bi-O bonds are less directional with the shrinking of the volume with the hole injection.

\begin{figure*}[htbp]
  \centering
  \includegraphics[width=0.85\textwidth]{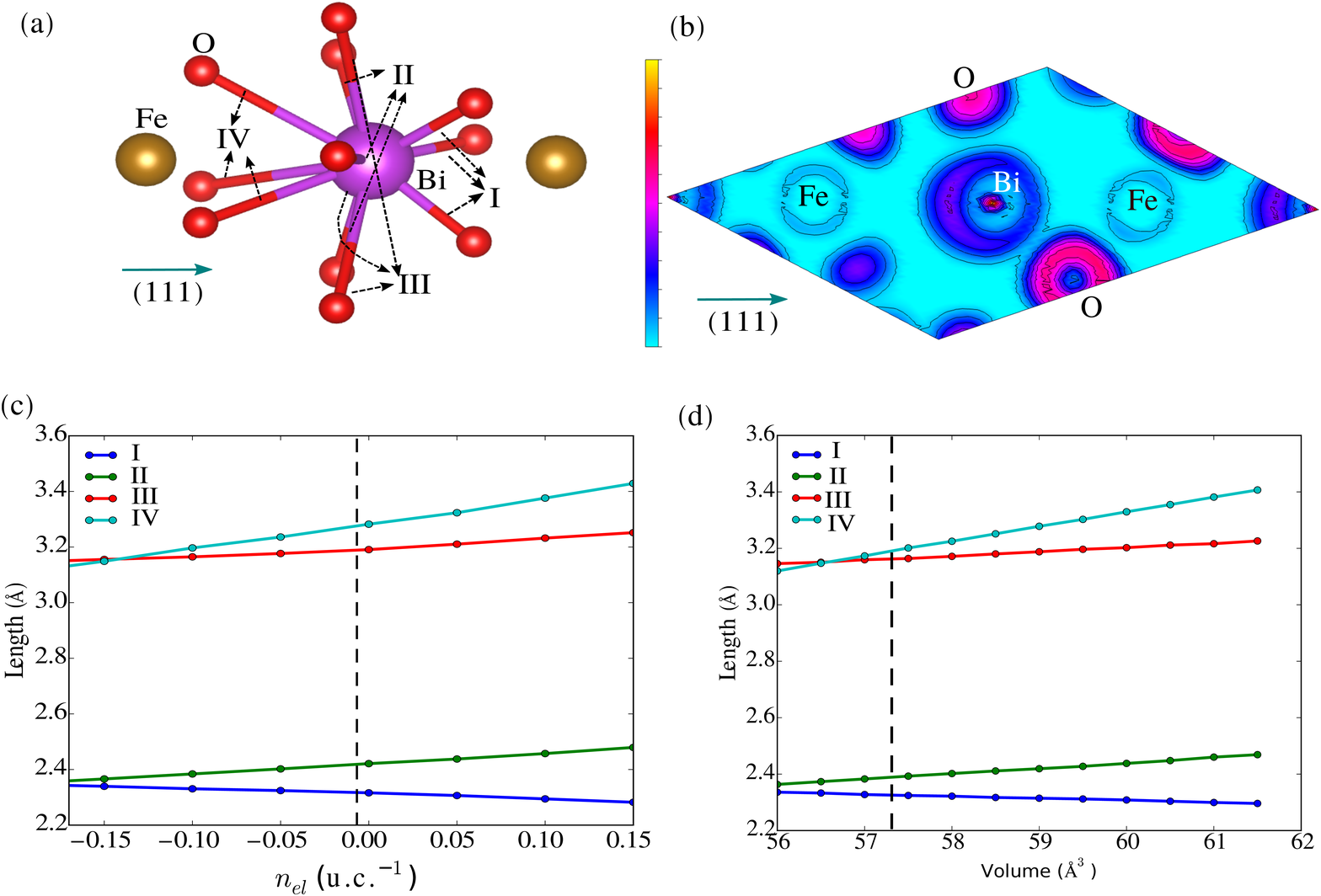}
  \caption{The directional Bi-O bonds and the Bi $6s$ lone pair. (a) The 12 Bi-O bonds of each Bi ion, which can be categorized into 4 groups labeled as I, II, III, and IV, respectively, because of the 3 fold rotation symmetry. (b) The contour map of the electron localization function in the cut of the diagonal plane of the R3c primitive cell. (c) The lengths of the Bi-O bonds versus the concentration of injected carriers. (d) The lengths of the Bi-O bonds versus the volume of the BFO unit with hydrostatic pressure. The black dashed lines present the point of the phase transition between $R3c$ and $Pbnm$.}
  \label{fig:lonepair}
\end{figure*}

To see whether above speculation is true or not, we analyzed the Bi-O bonds in BFO. With the electronic configuration of Bi$^{3+}$ ion being $6s^2p^0$, Bi ions can shift away from the central symmetric positions, forming Bi-O bonds on one side of Bi atoms and the lone pairs on the other side of Bi atoms\cite{PhysRevB.74.224412}. The forming of the lone pairs costs energy, while the forming of Bi-O covalent bondings gains energy. Therefore, if the Bi-O covalent bonding is strong enough, the forming of lone pairs and directional Bi-O bonds is stabilized, leading to the structural off-centering in BFO. In the $R3c$ structure with ferroelectric polarization in the [111] direction, Bi ions has 12 O neighbors. Because of the 3 fold rotation symmetry, these bonds can be divided into 4 groups labeled as I, II, III, and IV, as shown in Fig.~\ref{fig:lonepair} (a). The Bi-O bonds on the [111] direction side (group I) are shorter than those on the opposite side (group IV), leading to the Bi lone pair on the opposite to the polarization in BFO, which can be seen from the electron localization function\cite{ELF} in Fig.~\ref{fig:lonepair} (b). We compared the evolution of the Bi-O bond lengths with carrier concentration shown in Fig.~\ref{fig:lonepair} (c) to that with hydrostatic pressure in Fig.~\ref{fig:lonepair} (d), and found almost identical evolution patterns. The difference between the Bi-O bond lengths of group I and IV reduces with hole injection, which is the same with the hydrostatic pressure. Therefore, we can reach the conclusion that the hole injection leads to the reduction in volume and causes less directional Bi-O bonds and the weaker Bi lone pairs. Thus the structural off-centering is reduced. In the non-ferroelectric $Pbnm$ structure, the Bi-O bonds are less directional, which is compatible with the suppressing of the lone pair. Therefore, the non-ferroelectric $Pbnm$ phase is favored over the $R3c$ Phase.

%The hydrostatic pressure caused structural phase transition occurs when the volume of each BFO unit is compressed from the bulk value of 59 \AA$^3$ to a critical value of 57 \AA$^3$. Whereas the critical value of structure volume at phase transition with hole injection is 58 \AA$^3$, which is larger than that with hydrostatic pressure. The reason is that there is a driving force for the phase transition other than the volume shrinkage. This driving force comes from the weakening of the covalent component in the bonds of Bi with the oxygen atoms (in the bonds labled as I in Fig.~\ref{fig:lonepair} (c)) .The valence band maximum (VBM) are mainly consist of O $2p$ bands. Thus, by removing electrons from the VBM, the lengths of Bi-O bond increase due to the decreased Coulomb attraction between Bi cations and O anions, leading to the weaker covalent component of Bi-O bonding. As the off-centering of Bi atoms is stabilized by covalent component of the Bi-O bonding, it decreases with the injection of holes. Eventually, the energy cost to form the lone pairs surmount the energy gained from the Bi-O bonding, thus the structural off-centering is energetically unfavorable and the phase transition occurs.

The enhancement the structural off-centering with electron injection suggests that the screening of the long-range Coulomb interaction does not necessarily kill the the off-centering. This supports the idea that ferroelectric long-range order can be driven by short-range interactions\cite{PhysRevB.90.094108}. In the case of BFO, this short-range interaction is the cooperative shift of the Bi cations driven by the formation of lone pairs, which is not impaired by the screening of long-range Coulomb interaction. On the contrary, the free electrons on the CBM (mostly Fe $3d$ bands) pushes the surrounding oxygen anions away, reducing the lengths of the Bi-O bonds labeled as I. Thus the lone pair and the structural off-centering are strengthened.

\section{Conclusion}
In summary, we studied the electronic and lattice structure evolution of BFO with various concentrations of injected electrons and holes. We found that the electrons tend to be localized, which is stabilize by the electron-electron Coulomb repulsion and the expansion of the oxygen octahedron near the Fe site where the electron resides. The antiferromagnetic order also stabilize the localization. The injected holes tend to be delocalized if the O $2p$ on-site Coulomb interaction is weak (in other words, $U_\mathrm{O}$ is small). Small polarons are formed on O sites if $U_\mathrm{O}$ is large. With intermediate $U_\mathrm{O}$, multi-center polarons can be formed. The forming of hole polarons is also stabilized by the lattice distortion.

In the $R3c$ structure with injected carriers, delocalized electrons tend to enhance the off-centering, indicating that the ferroelectricity in BFO is not driven by long-range Coulomb interaction but the cooperative shift of Bi ions. Whereas holes tend to reduce the off-centering. With the hole concentration larger than $8.7\times10^{19} cm^{-3}$, there is a phase transition from $R3c$ structure to the non-ferroelectric $Pbnm$ structure. The reduction of off-centering and the phase transition in BFO are due to the shrinking of the lattice. These results indicate that the carrier injection can be an efficient way to control the ferroelectric distortion if the holes tend to be delocalized.

\begin{acknowledgments}
The work was supported by the National Basic Research Program of China (Grant Nos. 2014CB921001 and 2012CB921403), the National Natural Science Foundation of China (Grant Nos. 11474349, 11574365, and 11404380), and the Strategic Priority Research Program (B) of the Chinese Academy of Sciences (Grant No. XDB07030200).
\end{acknowledgments}
\bibliography{mybib.bib}
\end{document}